\documentclass[twocolumn,preprint]{emulateapj}
\usepackage{graphicx}

\usepackage{natbib}
\newcommand{\degree}{\ensuremath{^\circ}}

\begin{document}

\title{Radio Astrometry Of The Triple Systems Algol And UX Arietis}
\author{W. M. Peterson, R. L. Mutel}
\affil{Department of Physics and Astronomy, University of Iowa, Van Allen Hall, Iowa City, Iowa 52242, USA}
%\author{R. L. Mutel}
%\affil{Department of Physics and Astronomy, University of Iowa, Van Allen Hall, Iowa City, Iowa 52240, USA}
\author{J.-F. Lestrade}
\affil{C.N.R.S., Observatoire de Paris, F-75014 Paris, FR}
\author{M. G\"udel}
\affil{Department Astronomy, University of Vienna, Vienna, AT}
%\author{C. Lynch}
%\affil{Department of Physics and Astronomy, University of Iowa, Van Allen Hall, Iowa City, Iowa 52240, USA}
\author{W. M. Goss}
\affil{National Radio Astronomy Observatory, Socorro, NM, USA 87801}

\begin{abstract}
We have used multi-epoch long-baseline radio interferometry to determine the proper motion and orbital elements of Algol and UX Arietis, two radio-bright, close binary stellar systems with distant tertiary components. For Algol, we refine the proper motion and outer orbit solutions, confirming the recent result of \citet{Zavala:2010} that the inner orbit is retrograde. The radio centroid closely tracks the motion of the KIV secondary. In addition, the radio morphology varies from  double-lobed at low flux level to crescent-shaped during active periods. These results are most easily interpreted as synchrotron emission from a large, co-rotating meridional loop centered on the K-star. If this is correct, it provides a radio-optical frame tie candidate with an uncertainty $\pm0.5$ mas. For UX Arietis, we find a outer orbit  solution that accounts for previous VLBI observations of an acceleration term in the proper motion fit. The outer orbit solution is also consistent with previously published radial velocity curves and speckle observations of a third body. The derived tertiary mass, 0.75 solar masses, is consistent with the K1 main-sequence star detected spectroscopically. The inner orbit solution  favors radio emission from the active K0IV primary only. The radio morphology, consisting of a single, partially resolved emission region, may be associated with  the persistent polar spot observed using Doppler imaging.
\end{abstract}

\keywords{astrometry, stellar dynamics, binaries:close, radio continuum:stars}

\section{Introduction}
Multi-epoch phase-referenced radio interferometry is a powerful method to determine astrometric properties of radio-loud objects in  a variety of astrophysical settings. Very long baseline interferometry (VLBI), with baselines exceeding 10,000 km, can provide astrometric accuracies between 10 and 100 $\mu$as \citep{Pradel:2006}, allowing direct distance measurements to many galactic and extragalactic radio sources with unprecedented accuracy using trigonometric parallax \citep[e.g.,][]{Reid:2009, Deller:2009,  Dzib:2010}. Other astrometric VLBI applications include tests of general relativity \citep{ Fomalont:2009}, precise determination of Earth's inertial frame 
\citep{Petrov:2007,Petrov:2009}, and searches for extra-solar planets in late-type stellar systems \citep{Bower:2009}. 

In this paper, we report on astrometric VLBI studies of two well-studied active stellar systems, Algol and UX Arietis. Both systems contain a very active K sub-giant tidally locked in a short period binary system with a distant third companion. In addition, both are relatively close ($\leq50$~pc), so the angular scale of the inner orbit is several times the resolving beam of a global VLBI array. This allows tracking the motion of the active component within the inner binary orbit. By imaging the radio-loud component over many epochs, it is possible to determine all astrometric parameters (parallax, proper motion, inner and outer orbital elements) with high accuracy. In addition, the resulting maps provide a direct probe of the geometry of the coronal structure. Finally, if the solution is sufficiently accurate, the position of active star can be inferred on the radio image, allowing a tie between the optical reference frame and the radio ICRF with sub-mas accuracy \citep{Lestrade:1999, Bourda:2010}.

Algol (HD 19356, B8V+K2IV, 29 pc), the prototype of the eponymous Algol binary class, has been well-studied since its identification as an eclipsing binary system  more than two centuries ago  \citep{Goodricke:1783}. The cooler subgiant secondary fills its Roche lobe, causing episodic mass transfer onto an accretion disk surrounding the main sequence primary \citep{Richards:1993}. The subgiant is chromospherically and coronally active,  with frequent strong flares observed at radio \citep[e.g.,][]{Mutel:1998, Richards:2003},  ultraviolet \citep[e.g.,][]{Stern:1995}, and X-ray \citep[e.g.,][]{Ottmann:1996, Schmitt:1999} wavelengths. The close binary pair has an orbital period of 2.83 days and is tidally locked. A distant tertiary companion (F1V, P = 1.86 yr) is in an eccentric orbit oriented nearly perpendicular to the inner binary orbit. 

Although most of Algol's orbital parameters for both inner and outer orbits were well-determined spectroscopically decades ago  \citep{Hill:1971, Bachmann:1975, Stein:1977}, several parameters were either undetermined or ambiguous until more recent direct imaging became possible. The inner binary's orientation was first determined by \cite{Lestrade:1993} who observed the active K-star's radio emission at both quadratures using a global VLBI array. \cite{Pan:1993}, using optical interferometric data, determined that the orientation of the outer orbit was nearly perpendicular to the inner orbit, confirming the position angle inferred from speckle data \citep{Bonneau:1979}. 

Determination of the inclination of the inner orbital plane has proved more problematic. \cite{Csizmadia:2009} imaged Algol using both an optical interferometer and VLBI. They found that the inner orbit was prograde, i.e. that the line-of-sight component of the orbital angular momentum pointed toward the observer. However \cite{Zavala:2010} imaged both the inner and outer binaries at multiple epochs using the NPOI 6-element interferometer and found that the inner orbit is retrograde. In addition, they found that \citet{Pan:1993}'s determination of the outer orbit ascending node longitude was in error by 180$^{\circ}$. 
 
Interpreting the physical properties of a stellar radio source critically depends on accurately registering the radio morphology with its location within the stellar system. Algol was the first stellar system ever imaged using VLBI \citep{Clark:1975}. They found that during an exceptionally large radio flare ($S\sim$600 mJy), the source characteristic size was comparable to the overall angular size of the inner binary system. The inferred high brightness temperature and broad spectrum were consistent with gyrosynchrotron emission from mildly relativistic electrons in a coronal magnetic field. Subsequent VLBI images \citep{Mutel:1985, Lestrade:1988, Mutel:1998,Massi:2002, Csizmadia:2009} confirmed this basic picture, but the limited angular resolution and astrometric accuracy of these observations taken separately prevented an accurate registration with the stellar components. However, recently \cite{Peterson:2010} observed Algol at six epochs using a high-sensitivity global VLBI array at a higher frequency (15~GHz) than past observations. These data allowed much better sensitivity and angular resolution. They found that the radio structure consisted of a large coronal loop that was centered on the active K star, aligned along the inner binary  axis, and co-rotating. In this paper we combine this dataset with archival phased-referenced VLBI observations made since 1983. This combined dataset allows a global solution for all astrometric parameters.

UX Arietis (HD 21242, G5V+K0IV, 50 pc) is an active double-lined spectroscopic binary system.  The primary K subgiant shows activity at radio through UV wavelengths very similar to Algol \citep{Carlos:1971,Trigilio:1998,Mutel:1998, Torricelli:1998, Lang:1988,Buccino:2009, Ekmekci:2010}. There is also chromospheric activity associated with the secondary, possibly caused by episodic mass-transfer associated with Roche-lobe overflow from the primary \citep{Huenemoerder:1989, Gu:2002, AarumUlvaas:2003a, Ekmekci:2010}.  \cite{Vogt:1991} used Doppler imaging to detect a large, stable polar spot on the K primary, as well as transient spots at intermediate latitudes, which appear to be preferentially oriented toward the G companion \citep{Rosario:2008}. As with Algol, VLBI imaging of the radio corona of UX Arieties \citep{Mutel:1984, Beasley:2000, Massi:2002} shows structure comparable to the inner binary separation and a brightness temperature consistent with gyrosynchrotron emission. 

A third component has long been suspected in the UX Arietis system, based on both radial velocity anomalies 
\citep{Duemmler:2001,Glazunova:2008}, speckle observations of a distant third body \citep[e.g.,][]{Mcalister:1987, Hartkopf:2000, Balega:2006}, and apparent acceleration in proper motion studies \citep{Lestrade:1999,Boboltz:2003, Fey:2006}. 
In this paper, we combine new multi-epoch VLBI measurements with archival VLBI data to solve for all astrometric components including a tertiary component in a long-period orbit. 

\section{Observations}
Algol and UX Arietis  were observed with a global VLBI array at six and four epochs respectively between June 2008 and October 2009. The array comprised ten 25 m telescopes of the Very Long Baseline Array (VLBA)\footnote{The National Radio Astronomy Observatory is operated by Associated Universities Inc., under cooperative agreement with the National
Science Foundation.}, the 100m Green Bank Telescope (GBT), and the 100m Effelsberg telescope (EB). Observations were performed at 15.4~GHz in dual-polarization mode with a bandwidth of 128~MHz in each polarization. The array synthesized beamsize  ($\sim0.5$ mas ) was considerably smaller than the projected angular separation of the inner binary orbit of either system, so we were able to map motion within the inner binary. In addition, we used both previously published and archival VLBA observations to supplement our data, allowing highly accurate global astrometric solutions over a time internal spanning more than 25 years.

We used the `nodding' phase-referencing technique \citep{Lestrade:1990,Beasley:1995,Fomalont:1995}, switching rapidly between a compact extragalactic ICRF source with well-determined coordinates \citep{Ma:1998} and the target star. For the Algol observations, we used two phase calibrators: The angularly close primary phase calibrator
ICRF source J031301.9+41200  ($\Delta\theta=1.0\degree$) and a secondary calibrator, J031049.8+381453 ($\Delta\theta=2.8\degree$). We alternated between two-minute scans on Algol and one-minute scans on the primary phase calibrator, with additional one-minute scans of the secondary calibrator every hour. For the UX Arietis observations,  we alternated between scans of UX Arietis and ICRF source J033630.1+321829 (a.k.a. NRAO 140, $\Delta\theta=4.2\degree$) using  a 90-second cycling time.

In addition to these observations, we included two other radio astrometric measurements for these stars. First, we included several previously published radio interferometer positions \citep{Lestrade:1993, Lestrade:1999, Boboltz:2003, Fey:2006}. Second, a number of phase-referenced observations of both Algol and UX Arietis exist in the NRAO VLBA archive \citep{Sjouwerman:2004}. We calibrated and imaged these data and present the resulting positions in this paper for the first time. All VLBI data analyzed for this paper are summarized in Table~\ref{table:vlbi-obs}. 
\begin{deluxetable*}{ccrcrrl}
\tablecaption{VLBI Observing Log}
\tablewidth{6.5in}
\tablehead{
   \colhead{Exp. Code} & \colhead{Nr. epochs} & \colhead{Dates} & \colhead{Freq. (GHz)} & \colhead {Cal. Source} & \colhead{Array}  & \colhead{Ref.}  }
\startdata
\cutinhead{Algol}
BM044 &  1  & 1995.309	                 &	8.4	& 0313+412 & VLBA + Y27             & \cite{Mutel:1998}  \\
BM074  & 3  &1997.303 - 1997.308	   &	8.4	& 0313+412 & VLBA +Y27 + EB      &   \\
BM109 & 3 & 1999.042 - 1999.058    & 8.4   & 0313+412 & VLBA + Y27 + EB     &  \\
BM267 & 6 & 2008.265 - 2008.628    & 15.4 & 0313+412  & VLBA + GBT + EB    &  \citet{Peterson:2010}\\
\cutinhead{UX Arietis}
- 	& 9 & 1983.568 - 1994.405 & 8.4 & 0326+277 & (a)& \cite{Lestrade:1999} \\
BB032 & 1 & 1994.833 & 8.4  & 0326+277 & VLBA & \citet{Beasley:2000} \\
BB049 & 6 & 1995.875 - 1995.885 & 8.4 & 0326+277 &  VLBA & \\
BG097 & 1 & 2001.128 & 8.4 & 0326+277 & VLBA + EB & \\
BM140 & 4 & 2001.728 - 2001.739 & 8.4 & 0326+277 & VLBA + EB & \citet{Massi:2002} \\
BP157 & 4 & 2009.638 - 2009.797 & 15.4 & NRAO140 & VLBA + GBT + EB & This paper
\enddata
\label{table:vlbi-obs}
\tablenotetext{a}{VLBI arrays varied with epoch, see \cite{Lestrade:1999}}
\end{deluxetable*}

\section{Analysis}

\subsection{Data Calibration and Imaging}

% Look at Lestrade's papers for the calibration method of his data
VLBI visibilities from all epochs were calibrated and imaged using the NRAO AIPS software package \citep{Greisen:2003}, except for the 1983.5-1994.4 UX Arietis positions that were analysed using the SPRINT software package \citep[cf.][]{Lestrade:1999}. We followed standard VLBI amplitude and delay/rate calibration procedures. Geometric delays introduced by small errors in the predicted values of the Earth Orientation Parameters (EOP) were corrected by downloading tables of the measured EOP for the time of each observation. We also corrected  ionospheric and tropospheric delays --- these are discussed in section~\ref{sec:accuracy} below.

The phase calibrator sources used in the observations are compact extragalactic core-jet sources (e.g. \citet{Mutel:1998}). We made self-calibrated images of the calibrators (SNR $\sim$5000:1) which we used to recompute the fringe solution. In principle, the core position of a calibrator could shift with frequency. However, \citet{Fomalont:2011} observed four ICRF calibrators from 8-43~GHz and found the core positions are coincident to within 0.02 mas. Hence, we have assumed that the core centroids of the calibrators have a negligible shift from 8-15~GHz.

For the Algol observations, we also corrected for source motion within the observing period. Since the radio emission in Algol is associated with the K sub-giant \citep{Lestrade:1993}, it may move nearly 2 mas during a 12-hour observation, especially if the observation is centered near one of the eclipses. Images made without correcting for this motion would result in a smeared-out source along the direction of motion. Using the already well-determined orbital elements of the inner binary in Algol, we introduced a time-varying phase-tracking correction to the visibilities (using AIPS task CLCOR) equal to the K star's position offset in the inner binary orbit.

In order to determine radio centroid positions, we made Stokes I images using target visibilities that had been corrected from linearly interpolated delay-rate and phase solutions of calibrator scans bracketing each target scan. We then self-calibrated these images in order to improve image fidelity and better discern the source structure. If the source was unresolved, we fit a Gaussian to the centroid of the emission (AIPS task JMFIT) and used the Gaussian FWHM/2 as the position uncertainty estimate. For epochs with a resolved double structure we found the centroid of each radio lobe and used the midpoint between the lobes as the radio position for that epoch. In both cases, the position uncertainties were increased if the calculated position uncertainty due to tropospheric and ionospheric delays (section~\ref{sec:uncertainty}) exceeded the image-based estimate.

\subsection{Parameter Fitting}
We performed a global least-squares minimization of the orbital and astrometric parameters designated `variable' in Tables~\ref{table:algol-params} and \ref{table:uxari-params} using our VLBI data as well as radial velocity data \citep{Duemmler:2001, Hill:1971, Hill:1993, Gudel:1999} and differential positions from the Fourth Catalog of Interferometric Positions of Binary Stars \citep{Hartkopf:2010}. We began with a coarse grid search, varying each parameter by small steps over a reasonable range of values and finding the minimum $\chi^2$-value in the resulting multi-dimensional parameter space. We used the Nelder-Mead simplex (a.k.a. `amoeba') algorithm \citep{Press:1992} to refine the best-fit parameter values from the coarse grid search. This minimization scheme finds the direction of steepest descent without needing to calculate partial derivatives, which is especially useful when solving transcendental orbital equations. It also uses an adaptive stepsize in order to converge to within a very small tolerance of the minimum chi-square solution in the parameter space.

\subsection{Astrometric accuracy}
\label{sec:uncertainty}
\label{sec:accuracy}
An ideal interferometer with projected baseline length $B$ has a positional accuracy given by \citep{Thompson:1986}
\begin{equation}
\sigma = \frac{1}{2\pi}\ \frac{1}{SNR}\ \frac{\lambda}{B}\ ,
\end{equation}
where $SNR$ is the signal-to-noise ratio of the target source and $\lambda$ the observing wavelength. This ideal estimate, of order several $\mu$as for centimeter-wavelength VLBI, is almost never realized in practice since instrumental and intervening atmospheric effects degrade the phase of the incoming signals.

Phase-referenced astrometric observations using modern VLBI arrays typically have very small antenna position and timing uncertainties. Neither of these uncertainties contribute significantly to the astrometric accuracy so long as the switching time between calibrator and source is smaller than the coherence time. The two most important sources of position uncertainty result from path length fluctuations in the troposphere and the ionosphere \citep[e.g.][]{Fomalont:1995, Ros:2005,Pradel:2006}. A good phase-referencing scheme (i.e. with a short enough cycle time) should compensate for these delays, but for sub-mas astrometry they can still be a source of error and should be addressed.

The daytime ionospheric delay is dispersive ($\tau\propto\lambda^2$)  and is highly dependent on time of day. At 10~GHz, the daytime zenith ionospheric delay is quite large, typically 1~ns, corresponding to an excess path length 30~cm. Fortunately, since 1998 there have been several global databases of ionospheric delay maps available for online download. These maps are generated every two hours and are derived from observed GPS delays \citep{Ros:2000}. We have corrected  all post-1998 VLBI data for these delays (AIPS task TECOR).  \cite{Ros:2000} found that the residual ionospheric uncertainty after this correction is less than $\pm$0.15 ns (4.5 cm) at 8.4~GHz on global baselines. 

The tropospheric delay is non-dispersive, and consists of a dry component well predicted by the local atmospheric pressure, and a highly variable wet component whose value can exceed 10 cm excess path length. Both GPS delay maps and water vapor radiometers have been used to measure and correct for the wet component \citep[e.g.][]{Snajdrova:2006}, reducing the uncertainty to about 3 cm. However, although all VLBA sites measure the local atmospheric pressure and hence the dry component accurately, they are not equipped with water vapor radiometers, so there is no on-site measurement of wet component delay. Rather, the correlator applies a seasonally-based average value, which can be in error by several cm on a given day, especially for humid sites such as St. Croix. 

For phase-referenced observations on a single baseline, these path delay uncertainties result in a position uncertainty given by \citep[cf.][]{Reid:1999}

\begin{equation}\delta\theta = \left(\frac{c\ \delta\tau_0}{\lambda}\right){\rm tan}(Z)\ {\rm sec}(Z)\cdot \delta Z\cdot\Theta\ ,
\end{equation}
where $Z$ is the mean zenith angle, $\delta Z$ is the zenith angle difference between calibrator and target, $\delta\tau_0$ is the zenith path length uncertainty, and $\Theta$ is the baseline angular resolution. 

\begin{figure}
\centering
\includegraphics[width=3.5in]{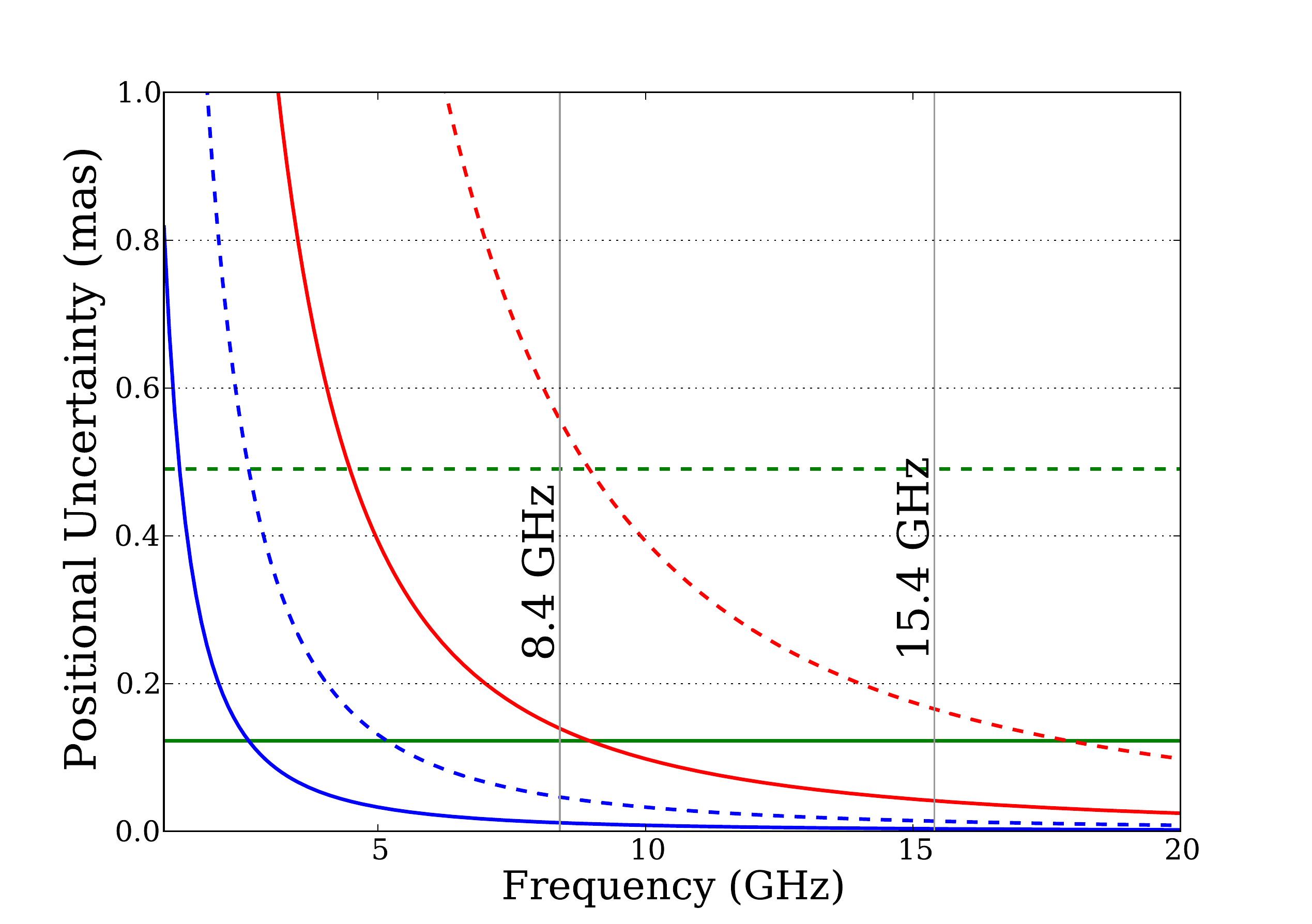}
\caption{Position uncertainty vs. frequency for baseline length $10^4$ km, 70\degree zenith angle,  tropospheric delay uncertainty $\pm$0.1 ns (3 cm) and daytime ionospheric delay uncertainty $\pm$0.1 ns (3 cm). The solid and dashed lines show position uncertainties for target-calibrator separations of 1\degree and 4\degree, respectively. Position uncertainties due to tropospheric delay uncertainty are shown by the green lines, the daytime ionosphere contribution is shown by red lines, and the nighttime ionosphere contribution is blue. The 15.4~GHz position uncertainties are dominated by the troposphere, while at 8.4~GHz, the daytime ionospheric contribution is comparable to the tropospheric contribution.}
\label{fig:delay-uncertainty}
\end{figure}

In Figure~\ref{fig:delay-uncertainty} we show a representative plot of the positional uncertainty expected from ionospheric and tropospheric delay uncertainties as a function of frequency for a global VLBI array using the nodding technique at mean zenith angle 70\degree and target-calibrator angular separations  1\degree and 4\degree. We have also assumed a tropospheric uncertainty $\pm$0.1 ns (3 cm), and mean daytime ionospheric uncertainty of $\pm$0.1 ns (3 cm). Nighttime ionospheric delay fluctuations are much smaller ($\sim$10\% daytime values) and are unimportant at frequencies above a few GHz. 

It is clear that at 15~GHz, the tropospheric contribution dominates, resulting in  position uncertainties 0.12 mas at 1\degree\ calibrator-target separation and 0.5 mas at 4\degree\ separation.  At 8.4~GHz, the daytime ionospheric and the tropospheric contributions are nearly equal, resulting in a total uncertainty  0.18 mas at 1\degree\ separation and  0.68 mas uncertainty at 4\degree\ separation. Unless given otherwise, we use these uncertainties as formal uncertainties for all VLBI measurements in Table~\ref{table:vlbi-obs}.
\begin{center}

\begin{deluxetable*}{c l l l  l  l l l l r r }
\tablecaption{Observed radio centroid positions}
\tablehead{
\colhead{JD} & \colhead {Mid-observation}  & \colhead{$\phi$} & \colhead {$\alpha$} & \colhead{$\delta$}  & 
 \colhead {$\alpha$} & \colhead{$\delta$}  & \colhead{$\sigma_{\alpha}$} & \colhead{$\sigma_{\delta}$} & \colhead{Freq.} & \colhead{Ref.}  \\
\colhead{2,400,000+} & \colhead{UT} & \colhead{Inner} & \colhead{geocentric} & \colhead{geocentric} \
& \colhead{heliocentric} & \colhead{heliocentric} & \colhead{mas} & \colhead{mas} & \colhead{GHz} & \colhead{} }
\startdata
\cutinhead{Algol}
47629.375 & 1989/04/12 21:00 & 0.28 & 03:08:10.127004 & 40:57:20.34098 & 03:08:10.125682 & 40:57:20.32477 & 0.75 & 1.50 & 8.4 & 1 \\
47632.302 & 1989/04/15 19:14 & 0.30 & 03:08:10.127162 & 40:57:20.34201 & 03:08:10.125973 & 40:57:20.32589 & 0.75 & 1.50 & 8.4 & 1 \\
47633.448 & 1989/04/16 22:45 & 0.70 & 03:08:10.126933 & 40:57:20.33808 & 03:08:10.125797 & 40:57:20.32200 & 0.75 & 1.50 & 8.4 & 1\\
47636.333 & 1989/04/19 19:59 & 0.71 & 03:08:10.127074 & 40:57:20.33958 & 03:08:10.126073 & 40:57:20.32364 & 0.75 & 1.50 & 8.4 & 1 \\
49831.333 & 1995/04/23 19:59 & 0.23 & 03:08:10.129554 & 40:57:20.33252 & 03:08:10.128722 & 40:57:20.31680 & 0.65 & 0.85 & 8.4 & 2 \\
50559.354 & 1997/04/20 20:30 & 0.13 & 03:08:10.130422 & 40:57:20.32512 & 03:08:10.129472 & 40:57:20.30925 & 1.15 & 1.55 & 8.4 & 2 \\
50560.354 & 1997/04/21 20:30 & 0.48 & 03:08:10.130524 & 40:57:20.32603 & 03:08:10.129621 & 40:57:20.31021 & 0.85 & 0.75 & 8.4 & 2 \\
50561.354 & 1997/04/22 20:30 & 0.83 & 03:08:10.130358 & 40:57:20.32413 & 03:08:10.129504 & 40:57:20.30838 & 1.00 & 1.15 & 8.4 & 2 \\
51194.583 & 1999/01/16 01:59 & 0.67 & 03:08:10.128550 & 40:57:20.34164 & 03:08:10.125901 & 40:57:20.33939 & 1.65 & 1.15 & 8.4 & 2 \\
51197.583 & 1999/01/19 01:59 & 0.72 & 03:08:10.128552 & 40:57:20.34124 & 03:08:10.125843 & 40:57:20.33815 & 0.60 & 0.75 & 8.4 & 2 \\
51200.583 & 1999/01/22 01:59 & 0.76 & 03:08:10.128433 & 40:57:20.33862 & 03:08:10.125672 & 40:57:20.33469 & 1.10 & 1.30 & 8.4 & 2 \\
51889.271 & 2000/12/10 18:29 & 0.95 & 03:08:10.130700 & 40:57:20.34500 & 03:08:10.129269 & 40:57:20.35259 & 1.00 & 7.00 & 8.4 & 3 \\
52212.000 & 2001/10/29 12:00 & 0.50 & 03:08:10.134300 & 40:57:20.33500 & 03:08:10.134977 & 40:57:20.35027 & 2.70 & 3.10 & 8.4 & 4 \\
54084.375 & 2006/12/14 21:00 & 0.50 & 03:08:10.133525 & 40:57:20.32636 & 03:08:10.131934 & 40:57:20.33302 & 3.00 & 3.00 & 5.0 & 5 \\
54563.333 & 2008/04/06 19:59 & 0.54 & 03:08:10.131456 & 40:57:20.32422 & 03:08:10.129887 & 40:57:20.30796 & 0.38 & 0.25 & 15.4 & 6 \\
54637.146 & 2008/06/19 15:29 & 0.29 & 03:08:10.136073 & 40:57:20.32786 & 03:08:10.137967 & 40:57:20.32258 & 0.53 & 0.47 & 15.4 & 6 \\
54652.042 & 2008/07/04 13:00 & 0.48 & 03:08:10.136676 & 40:57:20.32874 & 03:08:10.139083 & 40:57:20.32746 & 0.60 & 0.60 & 15.4 & 6 \\
54660.000 & 2008/07/12 12:00 & 0.26 & 03:08:10.137104 & 40:57:20.33069 & 03:08:10.139726 & 40:57:20.33159 & 0.55 & 0.60 & 15.4 & 6 \\
54675.000 & 2008/07/27 12:00 & 0.49 & 03:08:10.137484 & 40:57:20.33221 & 03:08:10.140380 & 40:57:20.33715 & 0.20 & 0.25 & 15.4 & 6 \\
54695.938 & 2008/08/17 10:30 & 0.79 & 03:08:10.137727 & 40:57:20.33316 & 03:08:10.140698 & 40:57:20.34317 & 0.20 & 0.25 & 15.4 & 6 \\
\cutinhead{UX Arietis}
45542.958 & 1983/07/27 11:00 & 0.96 & 03:26:35.337483 & 28:42:56.01878 & 03:26:35.338880 & 28:42:56.02219 & 1.00 & 1.00 & 8.4 & 7 \\
47228.250 & 1988/03/07 17:59 & 0.74 & 03:26:35.350239 & 28:42:55.53671 & 03:26:35.348903 & 28:42:55.53095 & 1.00 & 1.00 & 8.4 & 7 \\
48212.417 & 1990/11/16 21:59 & 0.61 & 03:26:35.360105 & 28:42:55.26662 & 03:26:35.360098 & 28:42:55.26999 & 1.00 & 1.00 & 8.4 & 7 \\
48420.208 & 1991/06/12 17:00 & 0.89 & 03:26:35.362638 & 28:42:55.20375 & 03:26:35.363319 & 28:42:55.20290 & 1.00 & 1.00 & 8.4 & 7 \\
48518.792 & 1991/09/19 07:00 & 0.20 & 03:26:35.364136 & 28:42:55.18201 & 03:26:35.365387 & 28:42:55.18796 & 1.00 & 1.00 & 8.4 & 7 \\
48520.792 & 1991/09/21 07:00 & 0.51 & 03:26:35.364132 & 28:42:55.18235 & 03:26:35.365355 & 28:42:55.18830 & 1.00 & 1.00 & 8.4 & 7 \\
48521.792 & 1991/09/22 07:00 & 0.67 & 03:26:35.363992 & 28:42:55.18042 & 03:26:35.365201 & 28:42:55.18638 & 1.00 & 1.00 & 8.4 & 7 \\
48636.792 & 1992/01/15 07:00 & 0.53 & 03:26:35.362634 & 28:42:55.14112 & 03:26:35.361381 & 28:42:55.13866 & 1.00 & 1.00 & 8.4 & 7 \\
48704.458 & 1992/03/22 23:00 & 0.04 & 03:26:35.363436 & 28:42:55.11757 & 03:26:35.362293 & 28:42:55.11167 & 1.00 & 1.00 & 8.4 & 7 \\
49501.292 & 1994/05/28 19:00 & 0.81 & 03:26:35.371336 & 28:42:54.89276 & 03:26:35.371675 & 28:42:54.89049 & 1.00 & 1.00 & 8.4 & 7 \\
49593.833 & 1994/08/29 07:59 & 0.19 & 03:26:35.373201 & 28:42:54.87460 & 03:26:35.374650 & 28:42:54.88009 & 1.10 & 1.19 & 8.4 & 8 \\
50037.854 & 1995/11/16 08:30 & 0.16 & 03:26:35.375143 & 28:42:54.74360 & 03:26:35.375156 & 28:42:54.74704 & 0.63 & 0.71 & 8.4 & 2 \\
50038.854 & 1995/11/17 08:30 & 0.31 & 03:26:35.374896 & 28:42:54.74219 & 03:26:35.374884 & 28:42:54.74554 & 1.42 & 1.45 & 8.4 & 2 \\
50039.854 & 1995/11/18 08:30 & 0.47 & 03:26:35.374982 & 28:42:54.74331 & 03:26:35.374944 & 28:42:54.74658 & 1.03 & 1.15 & 8.4 & 2 \\
50040.583 & 1995/11/19 01:59 & 0.58 & 03:26:35.374910 & 28:42:54.74244 & 03:26:35.374854 & 28:42:54.74565 & 1.09 & 0.87 & 8.4 & 2 \\
50041.781 & 1995/11/20 06:45 & 0.77 & 03:26:35.374954 & 28:42:54.74153 & 03:26:35.374867 & 28:42:54.74463 & 1.17 & 0.53 & 8.4 & 2 \\
50042.854 & 1995/11/21 08:30 & 0.93 & 03:26:35.375001 & 28:42:54.74069 & 03:26:35.374887 & 28:42:54.74369 & 1.55 & 0.88 & 8.4 & 2 \\
51889.271 & 2000/12/10 18:29 & 0.74 & 03:26:35.384900 & 28:42:54.17600 & 03:26:35.384290 & 28:42:54.17705 & 8.00 & 4.00 & 8.4 & 3 \\
51956.542 & 2001/02/16 01:00 & 0.19 & 03:26:35.384773 & 28:42:54.14555 & 03:26:35.383327 & 28:42:54.14058 & 0.85 & 1.11 & 8.4 & 2 \\
52175.792 & 2001/09/23 07:00 & 0.25 & 03:26:35.388388 & 28:42:54.08787 & 03:26:35.389573 & 28:42:54.09382 & 0.98 & 0.85 & 8.4 & 9 \\
52177.771 & 2001/09/25 06:29 & 0.55 & 03:26:35.388413 & 28:42:54.08608 & 03:26:35.389568 & 28:42:54.09202 & 1.17 & 0.39 & 8.4 & 9 \\
52178.792 & 2001/09/26 07:00 & 0.71 & 03:26:35.388498 & 28:42:54.08490 & 03:26:35.389636 & 28:42:54.09084 & 1.08 & 1.06 & 8.4 & 9 \\
52179.771 & 2001/09/27 06:29 & 0.86 & 03:26:35.388467 & 28:42:54.08532 & 03:26:35.389589 & 28:42:54.09125 & 0.69 & 0.83 & 8.4 & 9 \\
52212.000 & 2001/10/29 12:00 & 0.87 & 03:26:35.387700 & 28:42:54.08600 & 03:26:35.388147 & 28:42:54.09072 & 3.70 & 4.70 & 8.4 & 4 \\
55064.867 & 2009/08/21 08:48 & 0.01 & 03:26:35.412076 & 28:42:53.23876 & 03:26:35.413554 & 28:42:53.24391 & 0.45 & 0.28 & 15.4 & 6 \\
55089.850 & 2009/09/15 08:23 & 0.89 & 03:26:35.412153 & 28:42:53.23241 & 03:26:35.413447 & 28:42:53.23833 & 0.61 & 0.33 & 15.4 & 6 \\
55119.763 & 2009/10/15 06:18 & 0.54 & 03:26:35.411717 & 28:42:53.22360 & 03:26:35.412492 & 28:42:53.22906 & 0.48 & 0.28 & 15.4 & 6 \\
55122.721 & 2009/10/18 05:18 & 1.00 & 03:26:35.411888 & 28:42:53.22299 & 03:26:35.412597 & 28:42:53.22832 & 0.65 & 0.31 & 15.4 & 6 \\
 \enddata
 \tablecomments{(a) Heliocentric positions were calculated using the parallax given in Tables \ref{table:algol-params} and \ref{table:uxari-params}. (b) Coordinates were measured with respect to primary phase calibrators: 0313+412  (03:13:01.962129, +41:20:01.18353),   0326+277 (03:29:57.669425, +27:56:15.49901), NRAO140 (03:36:30.107609, +32:18:29.34226)  \citep{Ma:1998, Fey:2004}. (c) Previously published star coordinates have been corrected for different assumed calibrator positions if needed. (d) References: 1. \cite{Lestrade:1993}, 2. Positions from VLBA archival data, 3. \cite{Boboltz:2003}, 4. \cite{Fey:2006}, 5. \cite{Csizmadia:2009}, 6. This paper, 7. \cite{Lestrade:1984,Lestrade:1999} and unpublished positions, 8. \cite{Beasley:2000}, position calculated from VLBA archival data  9. \cite{Franciosini:1999}, positions calculated from VLBA archival data.}
 \label{table:vlb-coords}
\end{deluxetable*}
\end{center}

\section{Results}
Table~\ref{table:vlb-coords} lists radio centroid positions for both sources determined at all VLBI observing epochs. The global least-squares astrometric solutions computed from these positions are listed in Tables~\ref{table:algol-params} and \ref{table:uxari-params}. The corresponding differences between observed positions at each epoch and those calculated using the global solution are given in Table~\ref{table:fit-differences}, where the column labeled $\sigma$ is the number of standard deviations between the observed and model positions. For both solutions,  the agreement is excellent, with only a few epochs marginally above $2\sigma$. 
\begin{deluxetable*}{l r r l l}
\tablecaption{Algol orbital element solutions}
%\tablewidth{4in}
\tablehead{  \colhead{Parameter} & \colhead{Symbol} & \colhead {Value} & \colhead{unit} & \colhead{Param.type}  }
\startdata
% Insert table output here
               Parallax & $\Pi$ & 34.7 $\pm0.6$ & mas & Variable \\ 
        Primary mass & $m_A$ &  3.70  & $M_{\sun}$  & Fixed \\ 
     Secondary mass & $m_B$ &  0.79  &  $M_{\sun}$ & Fixed \\ 
       Tertiary mass & $m_C$ &  1.51 $\pm0.02$ &  $M_{\sun}$ & Calculated \\ 
      R.A. proper motion & $\mu_{\alpha \cos \delta}$ &  2.70 $\pm$  0.07 & mas yr$^{-1}$  & Variable\\ 
  Dec. proper motion & $\mu_{\delta}$ & -0.80 $\pm$  0.09 & mas yr$^{-1}$ & Variable \\ 
       Fiducial\tablenotemark{a} R.A. & $\alpha_0$ & 03:08:10.132409 $\pm$  0.7 & mas & Variable \\ 
Fiducial\tablenotemark{a} Declination & $\delta_0$ & 40:57:20.3353 $\pm$  0.6 &   mas & Variable \\ 
      Fiducial epoch & $JD_0$ & 2452212.02 & JD & Fixed \\ 
       & & 2001.82553 & year & \\ 
\cutinhead{Inner binary} 
              Period & $P_1$ & 2.867329  & days & Fixed\\ 
        Eccentricity & $e_1$ & 0.000  &   & Fixed \\ 
         Inclination & $i_1$ & 99 $\pm$ 17 & deg  & Variable\\ 
Longitude of ascending node & $\Omega_1$ & 48 $\pm$ 20 & deg & Variable\\ 
Lonigtude of periastron & $\omega_1$ & 270  & deg & Fixed \\ 
     Semi-major axis & $a_1$ &  2.26 & mas & Fixed \\ 
     Time periastron & $T_1$ & 2441773.49 & JD & Fixed \\ 
     & & 1973.24638 & year &  \\ 
\cutinhead{Outer binary} 
              Period & $P_2$ & 679.5 $\pm$  0.3 & days & Variable \\ 
& & 1.86  & years & \\
        Eccentricity & $e_2$ &  0.16 $\pm$  0.02 &  & Variable \\ 
         Inclination & $i_2$ & 85.5 $\pm$  1.4 & deg & Variable\\ 
Longitude of ascending node & $\Omega_2$ & 130.7 $\pm$  3.5 & deg & Variable\\ 
Longitude of periastron & $\omega_2$ & 312.0 $\pm$  1.4 & deg & Variable\\ 
     Semi-major axis & $a_2$ & 95.4 $\pm$  0.5 & mas & Variable \\ 
     Time periastron & $T_2$ & 2446930.0 $\pm$  3.2 & JD & Variable \\ 
   & & 1987.36551 & year & \\ 
     \enddata
 \tablenotetext{a}{Center of mass of triple system at fiducial epoch}
 \label{table:algol-params}
\end{deluxetable*}
\begin{deluxetable*}{l r r l l}
\tablecaption{UX Arietis orbital element solutions}
%\tablewidth{5in}
\tablehead{  \colhead{Parameter} & \colhead{Symbol} & \colhead {Value} & \colhead{unit}  & \colhead{Param. type} }
\startdata
            Parallax & $\Pi$ & 19.90 & mas  & Fixed\\ 
          Primary mass & $m_A$ &  1.10 & $M_{\sun}$ & Fixed  \\ 
      Secondary mass & $m_B$ &  0.95 & $M_{\sun}$ & Fixed \\       
      Tertiary mass & $m_C$ &  0.75$\pm$0.01  &$M_{\sun}$ & Calculated\\ 
  R.A. proper motion & $\mu_{\alpha \cos \delta}$ & 44.96 $\pm$ 0.13 & mas yr$^{-1}$ & Variable \\ 
  Dec. proper motion & $\mu_{\delta}$ & -102.33 $\pm$0.09 & mas yr$^{-1}$ & Variable \\ 
       Fiducial\tablenotemark{a} R.A. & $\alpha_0$ & 03:26:35.38386 $\pm$ 1.2 &  mas & Variable \\ 
Fiducial\tablenotemark{a} Declination & $\delta_0$ & 28:42:54.2755 $\pm$  0.8 &  mas & Variable\\ 
      Fiducial epoch & $JD_0$ & 2451544.5 & JD & Fixed \\
     & & 2000.0 & year & \\
\cutinhead{Inner binary} 
              Period & $P_1$ & 6.4378553  & days & Fixed \\ 
        Eccentricity & $e_1$ & 0.000 &   & Fixed \\ 
         Inclination & $i_1$ & 59.2 & deg & Fixed \\ 
Longitude of ascending node & $\Omega_1$ & 82 $\pm$ 34 & deg & Variable\\ 
Longitude of periastron & $\omega_1$ & 180.0 & deg  & Fixed \\ 
     Semi-major axis & $a_1$ &  1.71  & mas & Fixed \\ 
     Time periastron & $T_1$ & 2450642.00075  & JD & Fixed\\ 
& & 1997.52705 & year & \\
\cutinhead{Outer binary} 
              Period & $P_2$ & 40548.6 $\pm$ 70.2 & days & Variable \\ 
 & & 111.02 & years & \\
        Eccentricity & $e_2$ &  0.77 $\pm$ 0.01 &  & Variable \\ 
         Inclination & $i_2$ & 93.3 $\pm$ 0.6 & deg & Variable \\ 
Longitude of ascending node & $\Omega_2$ & 58.9 $\pm$ 0.5 & deg & Variable\\ 
Longitude of periastron & $\omega_2$ & 274.9 $\pm$ 0.8 & deg  & Variable\\ 
     Semi-major axis & $a_2$ & 648.0 $\pm$ 0.7 & mas & Variable \\ 
     Time periastron & $T_2$ & 2451664.9 $\pm$ 34.3 & JD  & Variable\\ 
     &  & 2000.32964 & year  & \\ 
\enddata
\tablenotetext{a}{Center of mass of triple system at fiducial epoch}
\label{table:uxari-params}
\end{deluxetable*}
\subsection{Algol}
\subsubsection{Outer orbit}
\begin{figure}
\centering
\includegraphics[width=3.5in]{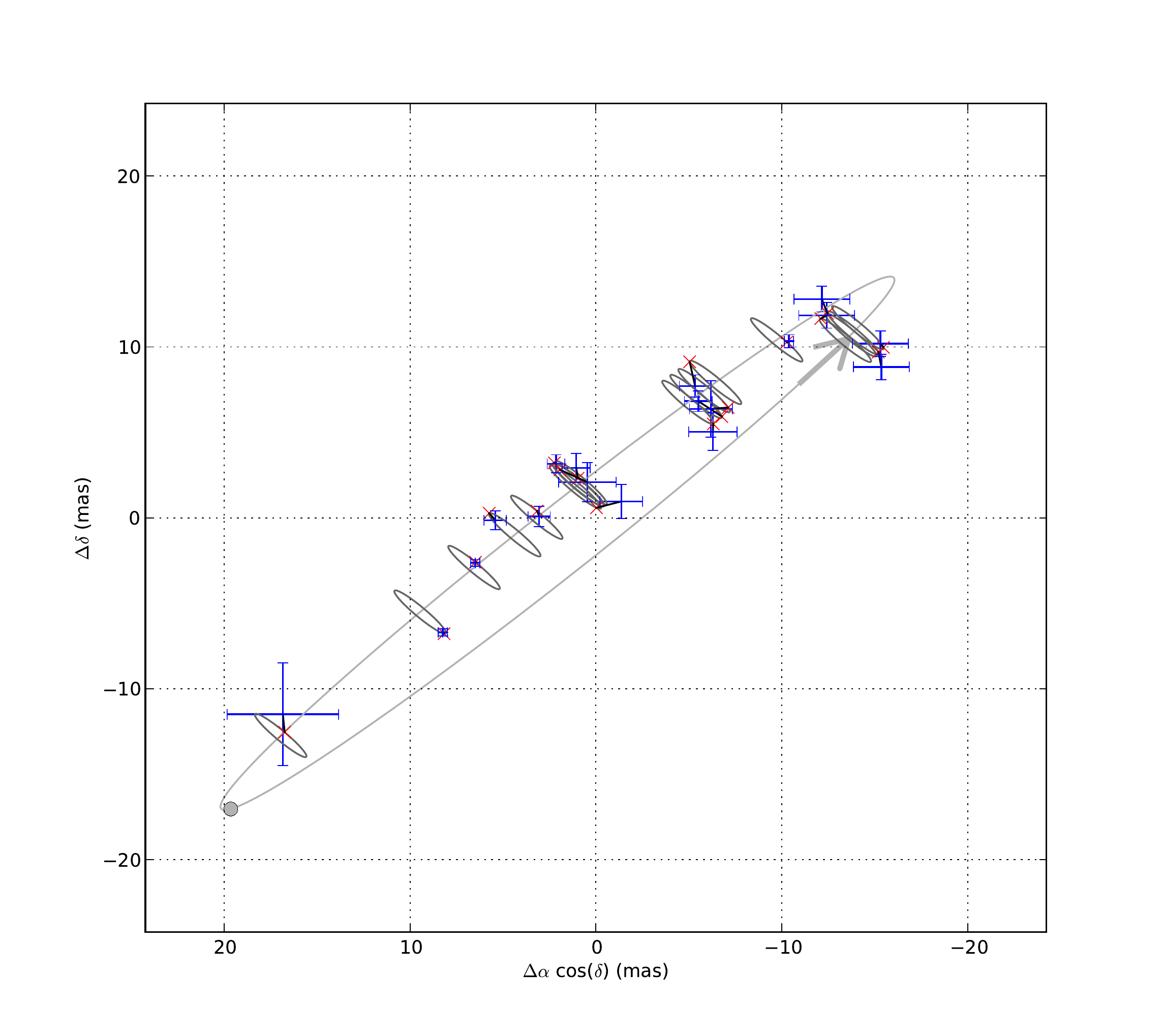}
\caption{Orbital path of the Algol inner binary center of mass (light gray line), with VLBI observations (blue errorbars) and model positions of the radio-loud star (Algol B, red $\times$'s). The model orbit of Algol B is shown at the time of each observation (dark gray ovals) for reference. The gray arrow shows the orbital direction for the AB center of mass, and the gray circle is the ascending node.}
\label{fig:algol-outer-orbit}
\end{figure}

Figure~\ref{fig:algol-outer-orbit} shows the position of the radio-loud secondary (Algol B) on the sky, after correction for proper motion and parallax. The model trajectory is shown as gray line, while the observed and predicted positions at each observing epoch are shown as blue errorbars and red $\times$'s, respectively. Our outer orbital solution agrees very well with \citet{Zavala:2010},  including the orientation of the Algol C orbit, which differed from earlier determinations (see Zavala et al. for discussion). We note that the derived mass of the tertiary component  ($1.57\pm0.01\ M_{\sun}$), although determined to high accuracy, depends on the masses of A and B, which we have assumed are exactly known. 

\begin{figure}
\centering
\includegraphics[width=3.5in]{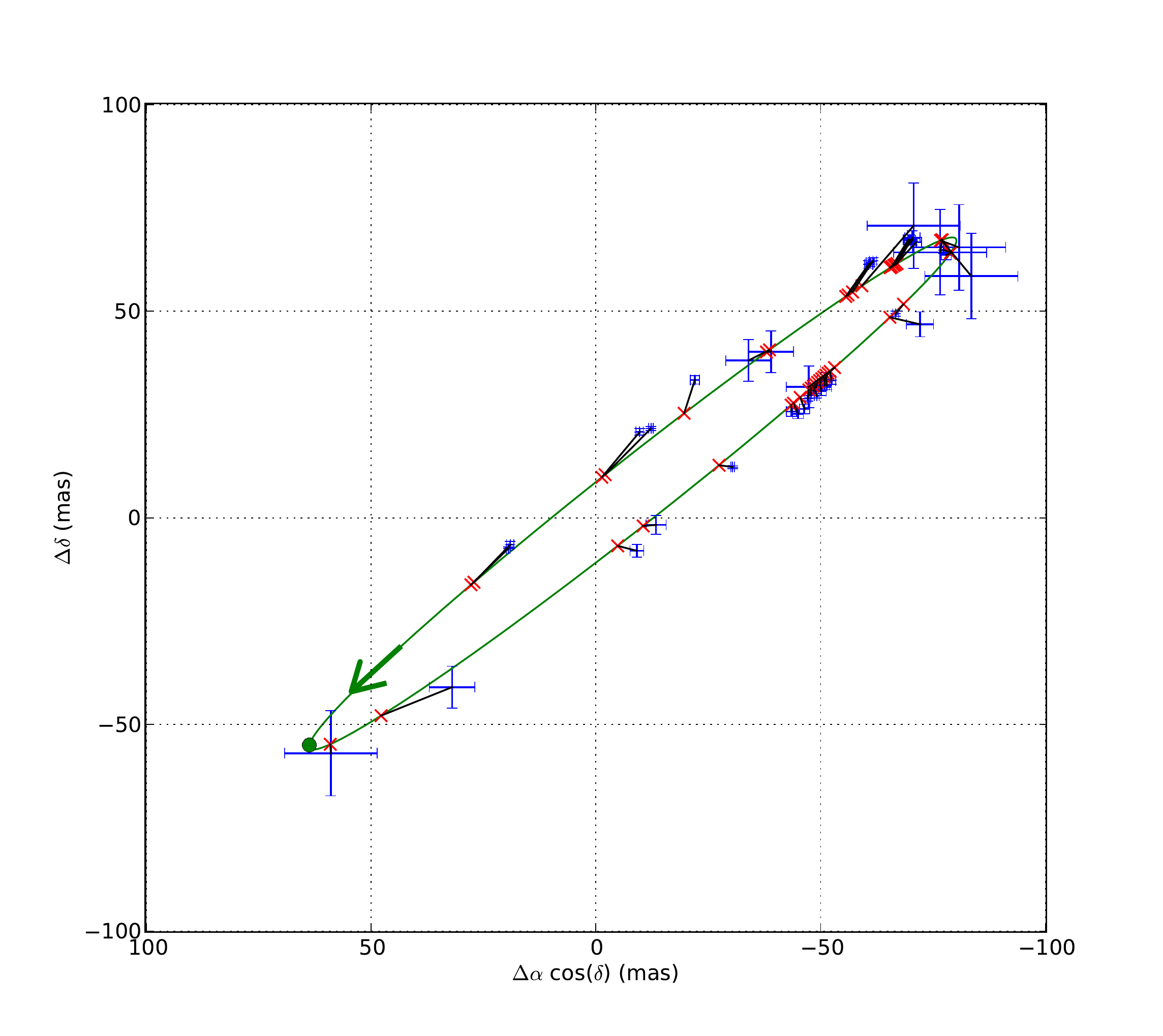}
\caption{Algol C (tertiary) orbit with respect to the AB center of mass determined from VLBI global astrometric solution (green line) along with speckle interferometer observations listed in the Fourth Catalog of Interferometric Measurements of Binary Stars  \citep{Hartkopf:2010} (blue errorbars) and model positions for the corresponding dates (red $\times$'s). The green arrow is positioned with its tail at the point of periastron, and shows the direction of orbit. The green circle is at the ascending nodal point. }
\label{fig:algol-speckle}
\end{figure}

We can use the outer orbit solution to compute the orbit of Algol C relative to the AB center of mass and compare with optical observations.  The model AB-C orbit is shown in Figure~\ref{fig:algol-speckle}, green line) along with optical observations of Algol C tabulated in the Fourth Catalog of Interferometric Measurements of Binary Stars (FCIMBS)  \citep{Hartkopf:2010}. The observed positions are in good agreement with the model orbit and predicted positions at each observing epoch.

\begin{centering}
\begin{deluxetable}{r r r r}
\tablecaption{VLBI --- model position differences}
\tablewidth{3.in}
\tablehead{\colhead{JD} & \colhead{$\Delta\alpha$} & \colhead {$\Delta\delta$} & \colhead{$\sigma$}  \\
 2,400,000+  &  mas  &  mas & mas }
\startdata 
\cutinhead{Algol}
47629.375  &   0.3  &  -0.2  &  0.4    \\
47632.302  &  -0.3  &  -0.8  &  0.7    \\
47633.448  &   0.2  &   0.9  &  0.6    \\
47636.333  &  -0.2  &  -0.2  &  0.3    \\
49831.333  &   0.3  &   1.4  &  1.7    \\
50559.354  &   1.5  &   0.7  &  1.4    \\
50560.354  &  -0.1  &  -0.6  &  0.8    \\
50561.354  &   1.3  &  -0.4  &  1.4    \\
51194.583  &  -0.9  &   0.1  &  0.6    \\
51197.583  &  -1.3  &  -0.9  &  2.4    \\
51200.583  &  -0.0  &   0.5  &  0.4    \\
%51889.271  &  -3.1  &   3.6  &  3.1    \\
%52212.000  &   0.8  &   1.7  &  0.6    \\
54084.375  &  -0.1  &  -1.1  &  0.4    \\
54563.333  &   0.1  &  -0.1  &  0.4    \\
54637.146  &   0.1  &   0.1  &  0.2    \\
54652.042  &   0.1  &   0.3  &  0.6    \\
54660.000  &   0.3  &   0.4  &  0.9    \\
54675.000  &  -0.0  &   0.0  &  0.2    \\
54695.938  &  -0.1  &  -0.1  &  0.5    \\
\cutinhead{UX Arietis}
45542.958  &   0.9  &  -1.5  &  1.7    \\
47228.250  &  -0.8  &  -0.2  &  0.8    \\
48212.417  &   0.7  &   0.9  &  1.1    \\
48420.208  &   1.1  &   0.8  &  1.4    \\
48518.792  &  -0.2  &   0.6  &  0.6    \\
48520.792  &  -1.4  &  -0.0  &  1.4    \\
48521.792  &   0.7  &   2.0  &  2.1    \\
48636.792  &  -1.3  &  -0.2  &  1.3    \\
48704.458  &  -1.3  &   0.7  &  1.5    \\
49501.292  &  -0.4  &   0.9  &  0.9    \\
49593.833  &  -0.6  &  -0.7  &  0.8    \\
50037.854  &  -0.5  &  -1.2  &  1.8    \\
50038.854  &   1.8  &  -0.3  &  1.3    \\
50039.854  &  -0.1  &  -1.5  &  1.3    \\
50040.583  &   0.7  &  -0.7  &  1.0    \\
50041.781  &   0.6  &   0.1  &  0.5    \\
50042.854  &   0.4  &   0.4  &  0.5    \\
50039.615  &  -1.6  &  -1.5  &  1.8    \\
50040.583  &   0.5  &  -0.6  &  1.2    \\
50041.615  &   0.4  &   0.7  &  1.3    \\
%51889.271  &   3.2  &  -3.6  &  1.0    \\
51956.542  &  -1.2  &  -1.2  &  1.7    \\
52175.792  &   1.3  &  -1.4  &  2.0    \\
52177.771  &  -0.2  &   0.2  &  0.6    \\
52178.792  &  -0.9  &   1.4  &  1.6    \\
52179.771  &   0.1  &   0.7  &  0.9    \\
%52212.000  &   3.7  &  -11.2  &  2.6    \\
55064.867  &   0.1  &   0.1  &  0.4    \\
55089.850  &  -0.6  &   0.4  &  1.5    \\
55119.763  &   0.6  &  -0.1  &  1.3    \\
55122.721  &  -0.6  &  -0.3  &  1.4    \\
 \enddata
\label{table:fit-differences}
\end{deluxetable}
\end{centering}

\subsubsection{Inner orbit}
\begin{figure}[t]
\centering
\includegraphics[width=3.5in]{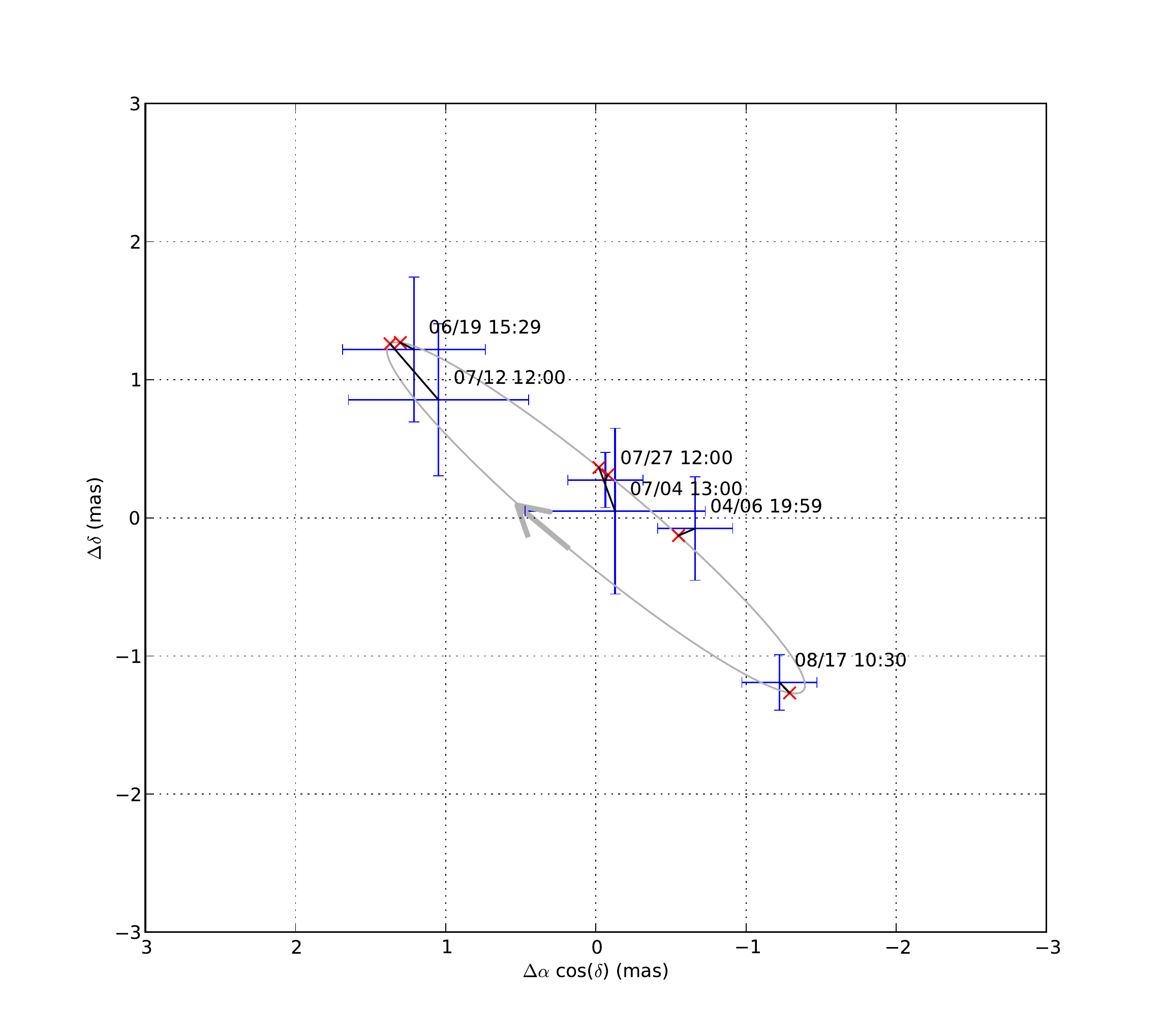}
\caption{Algol inner binary system, showing model orbit (gray line), and observed (blue errorbars) and predicted (red $\times$'s) radio centroids  at six epochs between 2008.27 --- 2008.63 . The gray arrow shows the direction of orbital travel. The ascending node of the orbit lies toward the upper-left in the figure.}
\label{fig:algol-inner-orbit}
\end{figure}

Figure~\ref{fig:algol-inner-orbit} shows the orbit of Algol B in the inner binary center of mass frame, along with the mean observed (blue errorbars) and model (red $\times$'s) positions at the six observing epochs observed at 15~GHz (experiment BM267). We have not displayed the 8~GHz positions, since their position uncertainty is twice as large as the 15~GHz points and provide little additional constraint of the inner binary orbit. The orientation of the inner binary, $\Omega=48\degree\pm20\degree$, agrees with the earlier VLBI result of \citet{Lestrade:1993} ($52\degree\pm5\degree$) and the more recent optical determination on \citet{Zavala:2010} ($47.4\degree\pm5.2\degree$). We also find $i= 99\degree\pm17\degree$, confirming the conclusion of \citet{Zavala:2010}, who found that the inner orbit is retrograde (i.e. $i>90\degree$, contrary to the previous determination of \citet{Csizmadia:2009}. 

\subsubsection{Algol's radio morphology: Co-rotating coronal loop centered on KIV star}
Algol's radio morphology consists of a double-lobed structure oriented normal to the inner orbital plane during quiescent states, and a crescent or loop structure during active periods \cite[see][and supplementary materials]{Peterson:2010}. The registration between radio and optical maps cannot be made directly, since the optical position uncertainty ellipse ($5 \times 8$ mas, \citet{Perryman:1997}) is larger than the inner binary orbit. However,  our 15.4~GHz observations show that the radio centroid mirrors the predicted motion of the KIV secondary over six epochs to an uncertainty  $\pm0.5$ mas.

It is possible that the radio centroid is systematically displaced from the K star and is following its trajectory precisely, but this seems unphysical. Occam's razor leads us to conclude that the correct registration is  that the radio centroid is coincident with the K star center. In that case,  the lobe structure is straddling the active KIV secondary, since the lobe separation ($\sim$1.0 mas) is the same as the K star's angular diameter (1.1 mas). 
As discussed by \citet[][and online supplement]{Peterson:2010}, the radio morphology and registration is consistent with a co-rotating, plasma-filled coronal loop structure emitting synchrotron radiation. The loop is oriented in the direction of the primary, which may imply  magnetic interaction with the primary's accretion disk  \citep[e.g.,][]{Retter:2005}.

X-ray observations \citep{Chung:2004} show that the centroid of the quiescent emission orbits with a semi-major axis 15\% smaller than the K star --- i.e. it is offset toward the center of mass of the inner binary. The radio positions in Figure~\ref{fig:algol-inner-orbit} appear likewise shifted inward of the K star orbit. Our global astrometric solution found that the semi-major axis of the centroid of the radio emission is shifted inward by 0.2 $\pm$ 0.8 mas. It is worth noting that the shift in the radio centroid is flux-dependent, moving inward toward the top of the co-rotating loop when the high-flux events fill the loop.

\subsection{UX Arietis}
\begin{figure}[b]
\centering
\includegraphics[width=3.5in]{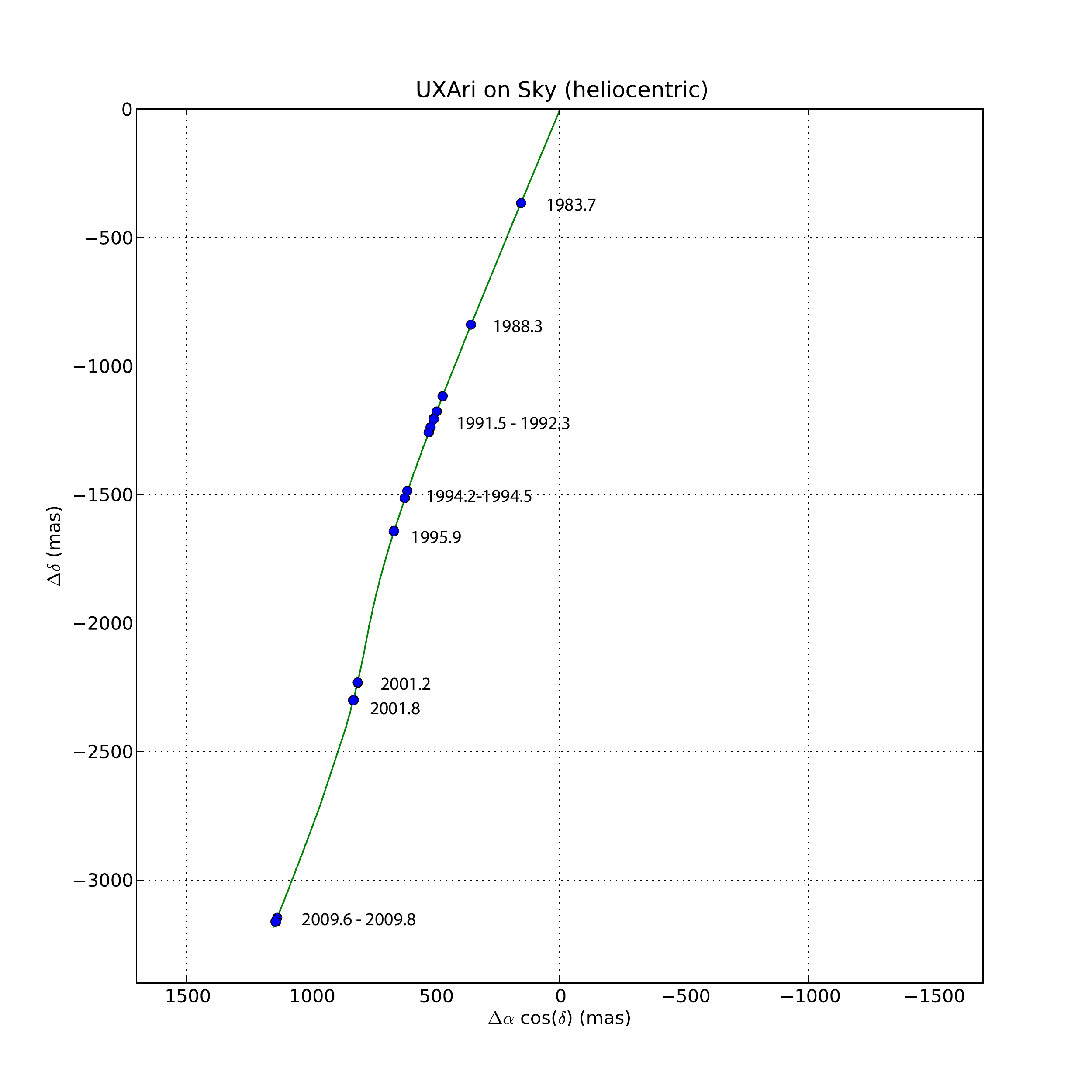}
\caption{UX Arietis radio position projected on the sky plane since 1980.0. The undulating trajectory results from proper motion combined with reflex motion in the AB-C outer binary.}
\label{fig:uxari-skyview}
\end{figure}
Unlike Algol, the orbital elements of a third component in the UX Arietis system were not previously known, although the presence of a third component is not unexpected --- companions in close binaries are quite common. \citet{Tokovinin:2006} found that 63\% of spectroscopic binaries (SB) that they surveyed had at least a tertiary companion. For SB's with a period less than 3 days, the fraction with companions is 96\%, suggesting that the shorter period systems exchanged angular momentum with their companions, shortening their orbital periods. 

In the UX Arietis system,  a third component had been inferred by detection of non-linear proper motion \citep[e.g., ][]{Boboltz:2003} and spectral lines of a possible third component \citep{AarumUlvaas:2003a}. However, the putative third component's orbit was poorly determined. The only published orbital solution consisted of two quite different models: a 10.7 yr circular orbit  and a 21.5 yr highly eccentric orbit  \citep{Duemmler:2001}. Therefore, it was necessary to undertake a comprehensive parameter search, constrained by the VLBI positions, third component radial velocity data  \citep{Duemmler:2001, AarumUlvaas:2003a, Glazunova:2008}, and optical interferometric observations in the FCIMBS \citep{Hartkopf:2010}. The search was complicated by the relatively high proper motion of UX Arietis ($\mu\sim$100 mas yr$^{-1}$), which magnifies a small fractional proper motion uncertainty over decades into a large aggregate position shift.

\subsubsection{Outer orbit}
\begin{figure}[b!]
\centering
\includegraphics[width=3.5in]{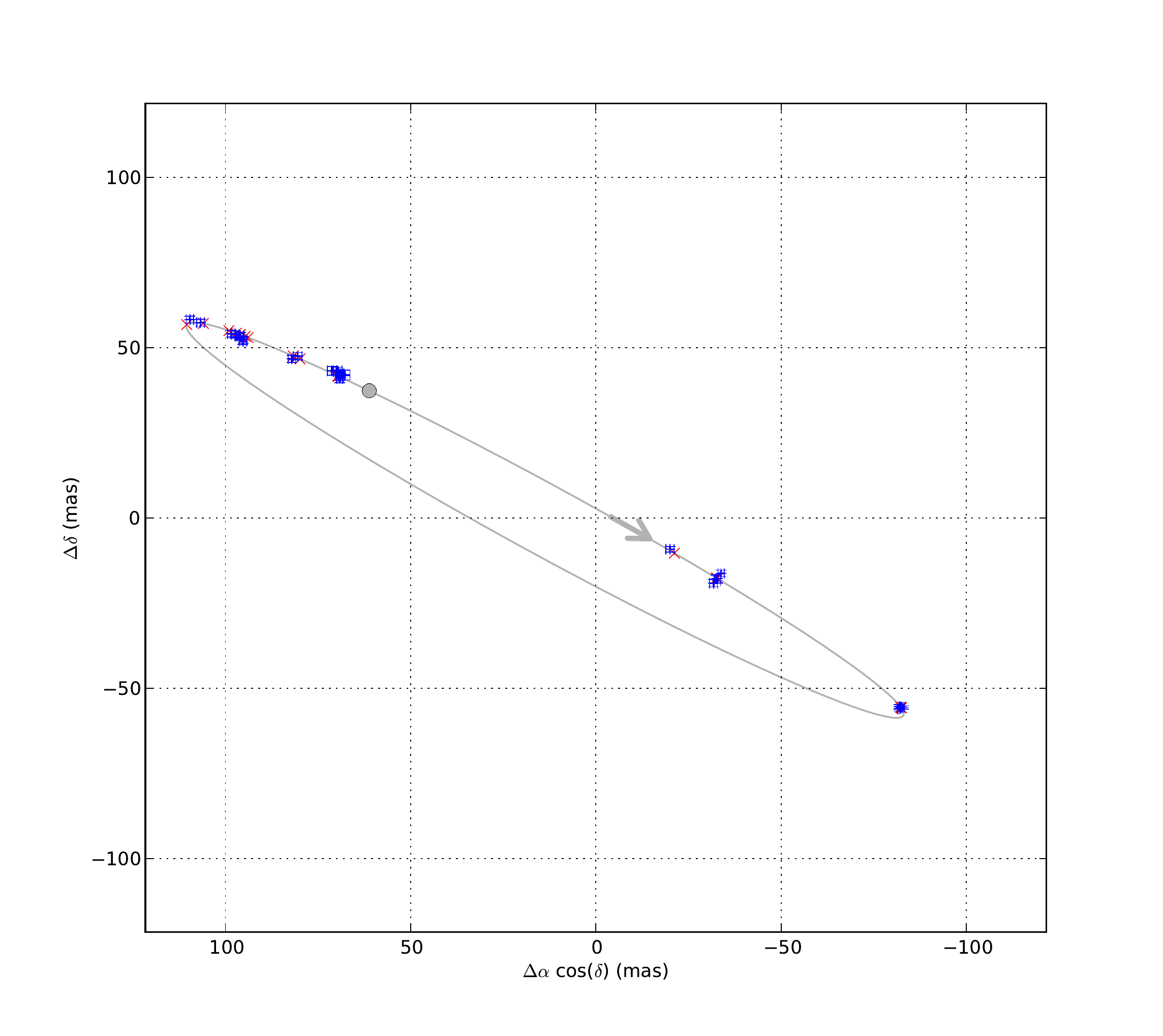}
\caption{Position of UX Arietis primary relative to triple system center of mass, showing model (gray line), and predicted (red~$\times$'s) and observed (blue~$+$'s) positions. The arrow shows the orbital direction and is positioned with its tail at the point of periastron. The gray circle shows the ascending nodal point.}
\label{fig:uxari-outer-orbit}
\end{figure}

Figure~\ref{fig:uxari-skyview} shows the radio position of UX Arietis on the sky plane, from epoch 1980.0 (origin) to 2009.8. The  blue circles are observed positions relative to 1985.0. The undulating shape of the trajectory, previously interpreted as an acceleration \citep{Lestrade:1999,Boboltz:2003,Fey:2006}) results from the reflex motion of the inner binary in a 111-yr period outer orbit. The position of the radio component in the frame of the outer binary center of mass in shown in Figure~\ref{fig:uxari-outer-orbit}. The gray line is the predicted position of the inner binary center of mass, with blue errorbars indicating observed positions and red $\times$'s the model positions after correction for proper motion and parallax. 

\begin{figure}
\centering
\includegraphics[width=3.5in]{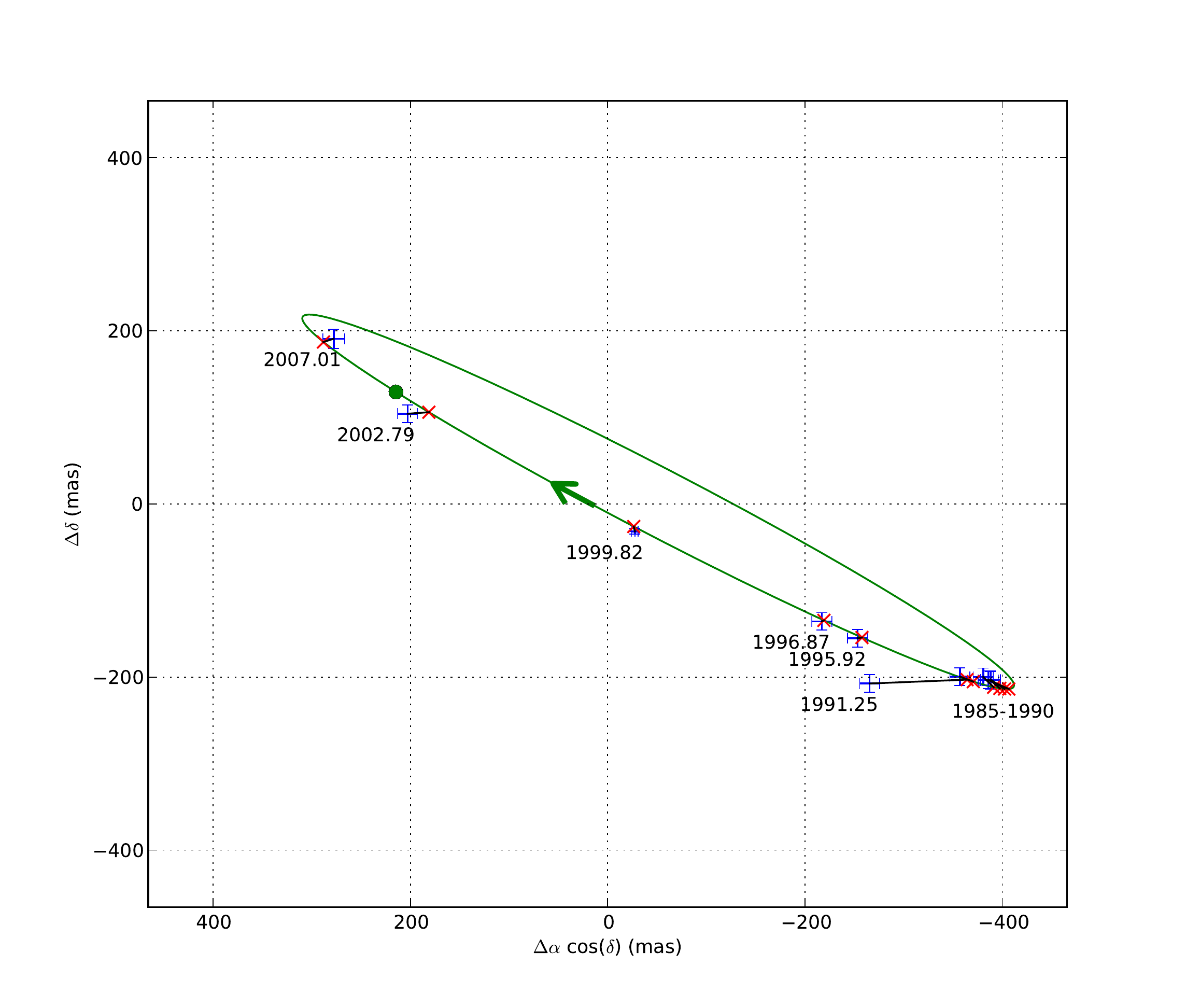}
\caption{UX Arietis C (tertiary) orbit with respect to the AB center of mass determined from VLBI global orbit solution (green line), along with speckle interferometer observations listed in the Fourth Catalog of Interferometric Measurements of Binary Stars  \citep{Hartkopf:2010} (blue errorbars) and model positions for the corresponding dates (red $\times$'s). The green arrow is positioned with its tail at the point of periastron, and shows the direction of orbit. The green circle is at the ascending nodal point. }
\label{fig:uxari-speckle}
\end{figure}

\begin{figure}[h]
\centering
\includegraphics[width=3.5in]{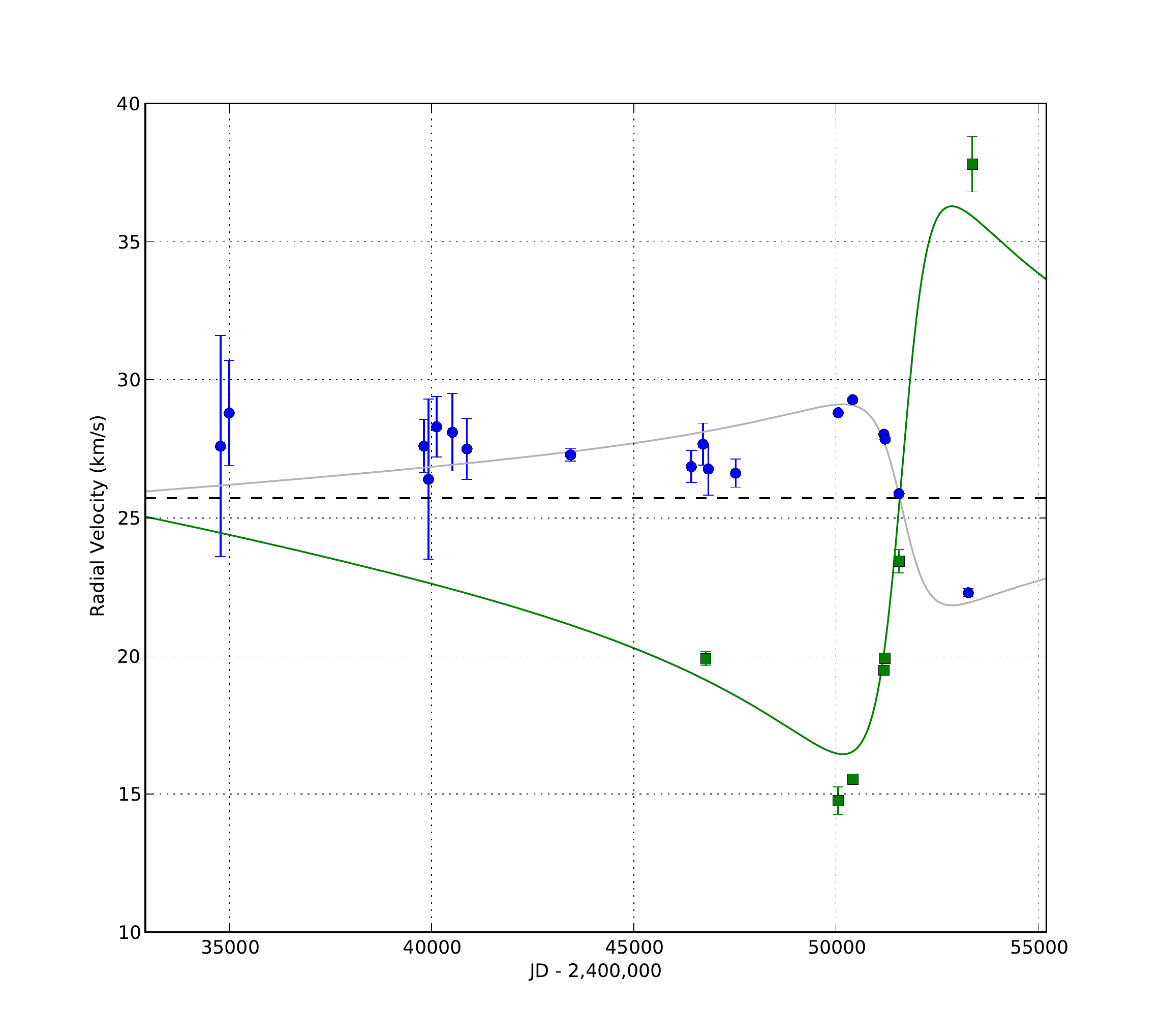}
\caption{UX Arietis outer orbit radial velocities \citep{Duemmler:2001,Massarotti:2008,Glazunova:2008}. Model curves based on the outer orbit solution are shown for the inner binary center of mass (gray line) and tertiary (green line), along with observed velocities for the tertiary (green squares) and AB center of mass (blue circles). }
\label{fig:uxari-outer-rv}
\end{figure}

We next compare the model solution with FCIMBS interferometer observations. Figure~\ref{fig:uxari-speckle} shows the model position of the tertiary component with respect to the inner binary center of mass (green line). The red and corresponding blue symbols are the model and observed positions of the tertiary component as listed in the FCIMBS.  Note that for all FCIMBS observations prior to 2002, we have applied a 180\degree correction to the catalogued position angle.  Finally, Figure~\ref{fig:uxari-outer-rv} shows the predicted radial velocity of the inner binary center of mass (gray line) and tertiary component (green line), along with corresponding radial velocity observations \citep{Duemmler:2001, Massarotti:2008, Glazunova:2008}. In general, the agreement with all three datasets is very good, although  the outlier at epoch 1991.25 from the HIPPARCOS catalog \citep{Perryman:1997} is not consistent with the orbit solution.

\subsubsection{Inner orbit}
%prrofread
Since both inner binary components of UX Arietis exhibit chromospheric activity \citep{AarumUlvaas:2003a}, it is unclear whether the radio emission originates from a single component, both components or perhaps an accretion region between the two stars. The orbital elements are well-characterized by spectroscopic observations except for the orientation ($\Omega$) and the inclination ($i\sim$60\degree, \citealt{Duemmler:2001}), which is degenerate about 90\degree reflection with respect to radial velocity curves. Hence we fixed the other orbital parameters and, as with Algol, used only the 15~GHz observations to solve for for $\Omega$ and $i$. 

We confirm that the K0IV primary is the source of the radio emission in UX Arietis by testing the quality of the fit assuming the alternatives: that the emission is either from the secondary, or from a location between the two components. The alternative assumptions both yielded unacceptable fits ($\chi^2_\nu = 7.6$ and $\chi^2_\nu = 2.8$, respectively), while the solution for radio emission from the primary was very good, yielding a good fit at the 99.7\% confidence level.

\begin{center}
\begin{figure*}[t]
\includegraphics[width=6.5in]{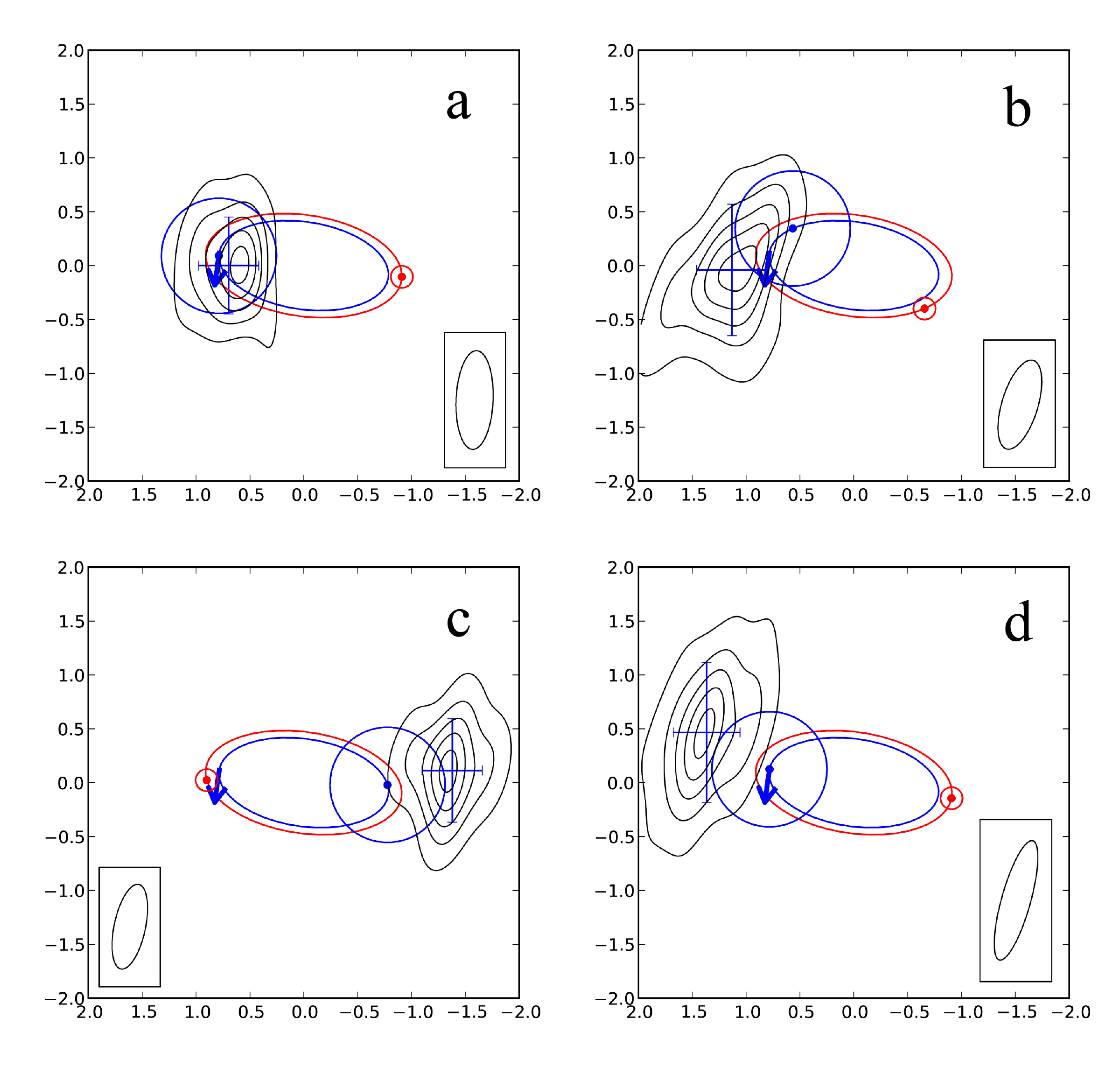}
\caption{UX Arietis inner orbit solution overlaid on radio contour maps at epochs (a) Aug 21, (b) Sep 15, (c) Oct 15, and (d) Oct 18, 2009. Orbits and predicted positions of the KIV primary (blue line, circle) and secondary (red line, circle) are shown. The blue arrow shows the orbital direction and position of the ascending node. The contour levels of the radio images are 10\%, 30\%, 50\%, 70\% and 90\%. of  peak values  0.7, 5.2, 3.8, and 1.4 mJy beam$^{-1}$ in panels a, b, c, and d respectively. The total flux density at each epoch is (a) 0.9 mJy, (b) 9.8 mJy, (c) 7.5 mJy and (d) 2.8 mJy.}
\label{fig:uxari-inner-orbit}
\end{figure*}
\end{center}

Figure~\ref{fig:uxari-inner-orbit} shows the best-fit orbital solution overlaid with contour maps of the radio emission at each epoch. We obtain a tentative solution for $\Omega$ ($82^\circ \pm 34^\circ$). Our data cannot constrain the solution in $i$, as can be seen by the fact that the timing of our observations were all very close to the nodal points in the inner binary orbit. Fixing the inclination at $59.2^\circ$ \citep{Duemmler:2001} produces a slightly better fit than the R.V.-degenerate value ($i=120.8^\circ$), thus we use the former value in our final solution.

\subsubsection{Source structure}
For all four VLBI observations made at 15~GHz, the self-calibrated radio images are amorphous, with an overall size comparable to the restoring beam ($\sim$0.4$\times$0.8~mas). This contrasts with earlier published images of UX Arietis \citep{Mutel:1985, Beasley:2000, Ros:2007} which show structure on an angular scale comparable to the binary separation (1.7~mas). However, the previous observations were at lower frequencies (5~GHz and 8.4~GHz) so it is possible that the interferometer phases were corrupted by uncorrected ionospheric delay fluctuations, especially for observations made prior to 1998, when GPS-based ionospheric corrections were unavailable (cf. section~\ref{sec:uncertainty}). Alternatively, there may be frequency-dependent extended structure with a steep spectral index that is not detectable at higher frequencies. 

\section{Discussion}

\subsection{Mass and spectral type of UX Arietis tertiary component}
The UX Arietis outer orbit solution  can be used to derive the mass of the tertiary component, $m_c = 0.75\pm0.01$~M$_{\sun}$, where the formal uncertainty is contingent on fixed values for the inner binary masses, as given in Table~\ref{table:uxari-params}.
This mass corresponds to a spectral type K1 main sequence star \citep{Zombeck:1990}. \citet{Duemmler:2001} estimated the mass of UX Arietis' tertiary component between $0.30 < m_3 < 0.46$~M$_\sun$, for their circular and elliptical orbit solutions respectively, although they reject the smaller mass, since it would be an M star too faint to be responsible for the spectral lines ascribed to the tertiary. \cite{AarumUlvaas:2003a} fit the continuum spectrum of the UX Arietis system and found that after correction for the primary and secondary spectra, the derived color index is best-fit to a K5 main-sequence star, although they cast doubt on this identiﬁcation, since the
implied radius (0.83 solar radii) is in the range of a K1-type star. Our derived mass agrees with the hotter type, so perhaps starspots on the tertiary are influencing its color index as is suspected with the primary \citep{AarumUlvaas:2003}.

\subsection{Comparison with polar loop models}

\cite{Peterson:2010} recently described a polar loop model for Algol in which the double-lobed radio components are associated with foot points of an active region of meridional magnetic field lines originating at the star's magnetic poles. A similar geometry was already suggested for this system by \cite{Franciosini:1999}. We now consider whether the single radio component seen in the UX Arietis system can be interpreted within this model.

A large, stable polar spot has been detected on the K primary of UX Arietis \citep{Vogt:1991,Elias:1995}.  These polar spots are a common phenomenon: Starspots on magnetically active cool stars preferentially appear near the poles \citep[e.g.][]{Hatzes:1996, Bruls:1998}. This preference for high latitudes may be due to rapid rotation, which leads to the Coriolis force dominating over buoyancy forces in the dynamics of magnetic flux tubes. As a consequence, flux tubes in the stellar convection zone migrate nearly parallel to the axis of rotation and thus surface at high latitudes \citep{Schuessler:1992}. Hence, magnetically active stars with rapid rotation exhibit magnetic flux eruption at high latitudes and polar starspots. 

The VLBI images of UX Arietis indicate that the radio centroid lies in the hemisphere facing away from the G star. This is consistent with observations of a similar late-type binary, HR1099, in which doppler imagery \citep{Donati:1992} and X-ray monitoring indicate that the active regions lie in the hemisphere facing away from the inactive companion. \citet{Audard:2001} suggest that this is because tidal locking of the rotation of the K star alters the internal dynamo in such a way that strong activity on the hemisphere facing the inactive star is suppressed.

Since the inner binary orbital plane is inclined at 30$^\circ$ to the observer's line of sight ($i\sim60$\degree), if the magnetic axis is close to perpendicular to the inner binary's orbital plane, only one pole will be visible. The polar loop model posits that the radio emission arises above both polar regions, but in this case, the far side emission will be largely occulted by the K star. This geometry may explain the absence of two radio centroids. 

\section{Summary}
We have analyzed more than 20 years of phase-referenced VLBI observations to determine accurate (sub-mas) positions of the radio centroids from two active close binaries, Algol and UX Arietis, both of which have distant tertiary companions. We used these positions to calculate proper motions and orbital elements of the inner and outer orbits in both stellar systems. 

For Algol, we confirm the early result of \citet{Lestrade:1993} that the radio centroid closely tracks the motion of the KIV secondary. Furthermore, the radio morphology, which varies from  double-lobed at low flux level to crescent-shaped during active periods, is consistent with synchrotron emission from a large, co-rotating meridional loop centered on the K-star. If this correct, it provides a radio-optical frame tie candidate with a precision $\pm$ 0.5 mas.  We also refine the proper motion and outer orbit solutions, confirming the recent result of \citet{Zavala:2010} that the inner orbit is retrograde.  

For UX Arietis, we find an tertiary orbit solution that accounts for previous VLBI observations of an acceleration term in the proper motion fit, as well as radial velocity curves and speckle observations. The dynamical mass, $0.75\pm0.01$ solar masses, supports the identification of the tertiary with a K1 main sequence star, consistent with third-body color index measurements  of \cite{AarumUlvaas:2003a}. The inner orbit solution  favors emission from the active K primary only, in the hemisphere facing away from the G star. The radio morphology, consisting of a single, partially resolved emission region, may be associated with the persistent polar spot observed using Doppler imaging.

\acknowledgements{The authors would like to thank Dr. Bob Zavala of the U.S. Naval Observatory Flagstaff Station for his helpful comments and insights, without which our paper would have been much less complete.}

\bibliography{rlm-library}

\clearpage
\clearpage

\end{document}